\RequirePackage{fix-cm}
\documentclass[twocolumn,epjc3]{svjour3}
\smartqed  
\usepackage{amsmath}
\DeclareMathOperator{\sech}{sech}
\usepackage{amssymb,bbold}
\usepackage{kantlipsum,widetext}
\usepackage{subcaption}

\RequirePackage{graphicx}
\RequirePackage[colorlinks,citecolor=blue,urlcolor=blue,linkcolor=blue]{hyperref}
\RequirePackage[numbers,sort&compress]{natbib}

\journalname{Eur. Phys. J. C}
\begin{document}

\title{Scattering, bound states, and resonances in the one-dimensional Dirac equation via supersymmetric quantum mechanics
}


\author{Camila C. Soares\thanksref{e1,addr1} \and Luis B. Castro\thanksref{e2,addr2,addr3} \and Antonio S. de Castro\thanksref{e3,addr4} 
}

\thankstext{e1}{e-mail: camila.soares@ifma.edu.br}
\thankstext{e2}{e-mail: lrb.castro@ufma.br}
\thankstext{e3}{e-mail: antonio.castro@unesp.br}


\institute{Departamento de Ensino, Instituto Federal do Maranh\~{a}o, Campus Z\'{e} Doca, 65365-000, Z\'{e} Doca, Maranh\~{a}o, Brazil. \label{addr1} \and
Coordenaç\~{a}o do Curso de F\'{i}sica - Bacharelado, Universidade Federal do Maranh\~{a}o, Campus Universit\'{a}rio do Bacanga, 65080-805, S\~{a}o Lu\'{\i}s, Maranh\~{a}o, Brazil. \label{addr2}\and
Programa de P\'{o}s-graduação em F\'{i}sica, Universidade Federal do Maranh\~{a}o, Campus Universit\'{a}rio do Bacanga, 65080-805, S\~{a}o Lu\'{\i}s, Maranh\~{a}o, Brazil. \label{addr3}\and
Departamento de F\'{\i}sica, Universidade Estadual Paulista, Campus de Guaratinguet\'{a}, 12516-410, Guaratinguet\'{a}, S\~{a}o Paulo, Brazil.\label{addr4}
}

\date{Received: date / Accepted: date}

\maketitle

\begin{abstract}

We develop a unified treatment of scattering and discrete spectra for the one-dimensional Dirac equation with scalar and vector interactions. Under the spin-symmetry condition, the coupled first-order Dirac system maps exactly onto an effective Sturm--Liouville (Schr\"o\-din\-ger-like) problem for a single spinor component. This mapping provides a convenient framework for analyzing transmission, reflection, and analytic continuation. As an explicit application, we consider effective interactions of hyperbolic P\"oschl--Teller type and exploit supersymmetric quantum mechanics and shape invariance to obtain a closed-form expression for the transmission probability. The bound-state spectrum is then recovered from the poles of the analytically continued transmission amplitude, reproducing known results and offering a unified description of scattering and bound states. For the barrier configuration, we briefly comment on the resulting pole pattern in the complex momentum plane and its connection with resonance and quasi-normal-mode behavior. Moreover, we use the chiral transformation to relate the spin- and pseudospin-symmetry sectors and translate results between them without repeating the full derivation.

\end{abstract}

\section{Introduction}
\label{intro}

Relativistic scattering in one spatial dimension provides a clean setting to investigate how
dissimilar Lorentz structures of external interactions shape both continuum and discrete spectra in the
Dirac theory. In particular, the one-dimensional Dirac equation with a Lorentz-scalar coupling
$S(x)$ and a time-like vector coupling $V(x)$ admits two especially important combinations,
$\Sigma(x)=V(x)+S(x)$ and $\Delta(x)=V(x)-S(x)$, which control the emergence of the so-called
spin and pseudospin symmetries. These symmetries, originally discussed in the context of
relativistic mean-field descriptions of nuclei and later explored more broadly, arise when one of
the combinations $\Delta$ or $\Sigma$ is (approximately) constant, leading to an SU(2)-like
structure that suppresses the corresponding spin-orbit splitting and simplifies the spectral
problem \cite{PRL78:436:1997,PR414:165:2005,PRA92:062137:2015}. In the present one-dimensional setting, we use the standard terminology to refer to the limits $\Delta(x)=0$ (spin symmetry) and $\Sigma(x)=0$ (pseudospin symmetry), even though the
usual $3+1$ dimensional interpretation is absent.

A key technical advantage of the spin- and pseudospin-symmetry limits is that the coupled
first-order Dirac system can be reduced \emph{exactly} to a single second-order
Schr\"odinger-like (Sturm--Liouville) equation for one spinor component, together with an
algebraic reconstruction formula for the other component. The price to pay is that the resulting
effective potential generally becomes energy dependent; nevertheless, the mapping allows us to
import a large set of nonrelativistic tools---including scattering theory and the analytic structure of transmission amplitudes---into the relativistic framework in a controlled way. Moreover, discrete transformations of the Dirac equation, notably the chiral transformation,
relate the spin and pseudospin sectors by exchanging $\Sigma\leftrightarrow\Delta$ while also
implementing $\psi_{+}\leftrightarrow\psi_{-}$ and $m\to -m$, enabling one to translate results
between the two symmetry limits without repeating the full derivation \cite{AP356:83:2015,PRA92:062137:2015}.

Among exactly solvable profiles, the hyperbolic P\"oschl--Teller (PT) potential plays a special role as a
benchmark model. In the relativistic context, PT-type scalar and vector couplings in the Dirac equation
(often under spin- and/or pseudospin-symmetry assumptions) have been investigated mainly from the bound-state
viewpoint by a variety of analytical schemes \cite{EPJA34:41:2007,ChPB20:070302:2011,JMP53:022104:2012}, and
related one-dimensional Dirac settings also display reflec\-tionless/PT--symmetric realizations
\cite{JPA43:075305:2010} and bound-state solutions \cite{EPL77:20009:2007,IJMPE16:3002:2007}. Outside the
Dirac setting, the PT barrier is a standard testbed for resonance pole patterns and quasi-normal-mode
constructions in different wave settings \cite{PRD30:295:1984,PLA380:1600:2016}. Recent examples where (modified) P\"oschl--Teller barriers are employed as effective or benchmark models for quasinormal-mode (QNM) spectra include near-extremal (A)dS black holes and related settings
\cite{CQG21:273:2003,PRD68:064007:2003,PRD81:125023:2010,PRD108:044032:2023,PRD110:064076:2024,JHEP2025:83:2025,PRD112:024031:2025}.

Supersymmetric quantum mechanics (SUSYQM) is intimately related to the factorization method, and the associated shape-invariance program provides a powerful algebraic route to exact spectra and, in many cases, to closed-form scattering data \cite{JETPL38:356:1983,JPA21:L501:1988,PR251:267:1995,JUNKER1996,BAGCHI2000,DONG2007}. SUSY and factorization techniques have also been applied to relativistic Dirac problems with scalar, vector and tensor interactions, including spin- and pseudospin-symmetric settings \cite{AP325:1720:2010,AP325:2522:2010}. In SUSYQM, Hamiltonians are organized into partner pairs sharing closely related spectral properties, and shape invariance generates solvable hierarchies that often permit compact expressions for reflection and transmission amplitudes. These methods are particularly natural for the P\"oschl--Teller family and its supersymmetric extensions \cite{ZP83:143:1933,FLUGGE1999,AJP75:1151:2007,JPA32:8447:1999}. In this case, shape invariance leads to compact analytic expressions for transmission and reflection probabilities. In the barrier regime, analytic continuation provides a convenient pole classification in the complex momentum plane
\cite{PLA380:1600:2016}, a feature that also underlies PT-based QNM constructions \cite{PRD30:295:1984}.

In this work, we develop a unified treatment of scattering and discrete states for the
one-dimensional Dirac equation with scalar and time-like vector interactions by exploiting the
exact mapping (under spin symmetry condition) to a Schr\"odinger-like problem for a
single spinor component. This is first carried out at a general level, emphasizing how the
relativistic dynamics is encoded into an effective Sturm--Liouville problem. As an explicit and
physically transparent application, we specialize to the hyperbolic P\"oschl--Teller profile,
for which SUSYQM and shape invariance allow us to obtain closed-form expressions for the
transmission amplitude. Bound states are then recovered from the
poles of the analytically continued transmission amplitude, reproducing known Dirac spectra for this profile
as a nontrivial consistency check \cite{EPL77:20009:2007}. Finally, we comment on how the pole structure
in the complex momentum plane encodes resonant states and QNMs when the effective interaction
takes a barrier form, connecting with standard PT-based QNM constructions \cite{PRD30:295:1984,PLA380:1600:2016}.

This paper is organized as follows. In Sec.~\ref{sec:ode}, we review the one-dimensional Dirac equation with
scalar and vector couplings, define the spin-symmetry limit, and present the
general reduction to a Schr\"odinger-like equation. In Sec.~\ref{sec:revSUSYQM}, we summarize the SUSYQM and
shape-invariance machinery needed for scattering. In Sec.~\ref{sec:PT_application}, we apply the formalism to the
P\"oschl--Teller family, deriving closed expressions for the transmission amplitude, discussing bound-state
poles and, for the barrier case, resonance/QNM pole families, and showing how the corresponding results for the pseudospin-symmetry sector ($\Sigma=0$) follow from the chiral transformation. Our conclusions are given in
Sec.~\ref{sec:fr}.
              
\section{The one-dimensional Dirac equation}
\label{sec:ode}

In this section we introduce the one-dimensional Dirac equation with scalar and vector interactions.
Our goal is to map the relativistic problem into a Schr\"odinger-like equation for one spinor component
under the symmetry condition $\Delta=0$. 

The one-dimensional time-independent Dirac equation for a fermion of
rest mass $m$ and a general potential $\mathcal{V}$ reads
\begin{equation}
H\Phi=E\Phi \label{eq1}\, ,
\end{equation}
\noindent with
\begin{equation}
H=c\alpha p+\beta mc^{2}+\mathcal{V}\, , \label{eq1a}
\end{equation}
\noindent where $E$ is the energy of the fermion, $c$ is the velocity of
light and $p$ is the momentum operator. $\alpha $ and $\beta $ are Hermitian
square matrices satisfying the relations $\alpha ^{2}=\beta ^{2}=1$, $%
\left\{ \alpha ,\beta \right\} =0$. The positive definite function $%
|\Phi |^{2}=\Phi ^{\dagger }\Phi $, satisfying a continuity
equation, is interpreted as a position probability density and its
integral over the whole space is a constant of motion. This interpretation is completely
satisfactory for single-particle states \cite{THALLER1992}. Also a well-known uniform charge current density can be expressed as
\begin{equation}\label{currentden}
J=c\,\Phi^{\dagger}\alpha\Phi\,.
\end{equation}
\noindent We set $\mathcal{V}$ to be
\begin{equation}
\mathcal{V}=IV_{t}+\beta V_{s}+\alpha V_{sp}\, . \label{eq2}
\end{equation}
\noindent The subscripts for the terms of
the potential denote their properties under a Lorentz
transformation: \textit{t} and \textit{sp} for the time and space
components of the two-vector potential, and \textit{s} for
the scalar term. To have an explicit
expression for the $\alpha$ and $\beta$ matrices one can choose
$2\times 2$ Pauli matrices that satisfy the same algebra. We use
$\beta=\sigma_{3}$ and $\alpha=\sigma_{1}$.

The Hamiltonian is invariant under the parity
operation, i.e., when $x\rightarrow -x$, if $V_{sp}(x)$ changes sign, whereas $V_{t}(x)$ and $V_{s}(x)$ remain the same. This
is because the parity operator is $P=\mathrm{exp} (i\eta
)P_{0}\sigma _{3}$, where $\eta$ is a constant phase and $P_{0}$
changes $x$ into $-x$. Because this unitary operator anticommutes
with $\alpha$, it changes sign under a parity
transformation, whereas $I$ and $\beta$, which commute with $P$,
remain the same. When one writes down the explicit equations of
motion in terms of the components of the spinor $\Phi$, the
combinations $\Sigma=V_{t}+V_{s}$ and $\Delta=V_{t}-V_{s}$ of the
vector and scalar components arise naturally. Therefore, it is
convenient to rewrite the Hamiltonian in terms of these
potentials. We have
\begin{equation}
H=c\alpha \left(p+\frac{ V_{sp}}{c} \right)+ \beta
mc^{2}+\frac{I+\beta }{2}\Sigma +\frac{I-\beta }{2}\Delta\, . \label{eq2b}
\end{equation}

\subsection{Chiral transformation}
\label{sec:sub:ct}

In the present work the discrete symmetry most relevant to relate the spin/pseudospin sectors is the chiral
transformation, since it provides a direct mapping between the limits $\Delta=0$ and $\Sigma=0$.
For this reason we focus on the chiral transformation below and use it later to translate scattering
results between the two symmetry sectors.

The chiral operator is the matrix $\gamma^{5}=\sigma_{1}$,
and we will call \textquotedblleft chiral
transformation\textquotedblright\ the transformation associated with
it. Thus, the transformed spinor is given by $\Phi_{\chi
}=\gamma^{5}\Phi$ and the transformed Hamiltonian $H_{\chi
}=\gamma^{5}H\gamma^{5}$. Because $\gamma^{5}$ anticommutes with
$\beta$, the time-independent chiral transformed Dirac equation is
\begin{equation}
H_{\chi }\Phi_{\chi }=E \Phi_{\chi} ,\label{eq6}
\end{equation}
\noindent where $H_{\chi }$ is given by
\begin{equation}\label{eq7}
H_{\chi }=c\alpha \left(p+\frac{ V_{sp}}{c} \right)-\beta
mc^{2}+IV_{t}-\beta V_{s}\,, 
\end{equation}
\noindent or
\begin{equation}\label{eq8}
H_{\chi }=c\alpha \left(p+\frac{ V_{sp}}{c} \right)-\beta
mc^{2}+\frac{I+\beta }{2}\,\Delta +\frac{I-\beta }{2}\,\Sigma\,.
\end{equation}
\noindent This means that the chiral
transformation changes the sign of the mass and of the scalar potential, thus turning $\Sigma$ into $\Delta$ and
vice versa. A chiral invariant Hamiltonian needs to have zero mass
and $V_{s}$ zero everywhere.

\subsection{Equations of motion}
\label{sec:sub:eom}

The space component of the two-vector potential in Eq. (\ref{eq2})
can be absorbed into the wave function by defining a new spinor
$\psi$ such that
\begin{equation}
 \Phi =e^{-i \Lambda} \phi \, , \label{eq9}
\end{equation}
\noindent in which $\Lambda(x) =(1/\hbar c)\int^{x}V_{sp}(y)dy$,
because we have $H\phi=e^{-i \Lambda}\left( H-\alpha V_{sp}
\right)\phi$. From this point on, we will refer to $V_{t}$ as simply
a vector potential, following the common usage of this term (usually
denoted by $V_{v}$). Writing the spinor $\phi$ in terms of
its components as
\begin{equation}
\phi =\left( \begin{array}{c} \phi_{+} \\ \phi_{-} \end{array}
\right) \, , \label{eq10}
\end{equation}
\noindent the Dirac equation yields two coupled first-order
equations:
\begin{eqnarray}
&&-i \hbar c \phi^{\prime}_{-}+mc^{2}\phi_{+}+\Sigma
\phi_{+}=E \phi_{+}\,,\label{eq11a}\\
&&-i \hbar c \phi^{\prime}_{+}-mc^{2}\phi_{-}+\Delta
\phi_{-}=E \phi_{-}\, ,\label{eq11b}
\end{eqnarray}
\noindent where the prime denotes differentiation with respect to
$x$. The normalization condition $\int^{+\infty }_{-\infty }dx(|\phi_{+}|^{2}+|\phi_{-}|^{2})=1$ implies that both $\phi_{+}$ and $\phi_{-}$ are square-integrable functions. Furthermore, Eqs.~(\ref{eq11a}) and (\ref{eq11b}) show that $\phi_{+}$ and
$\phi_{-}$ have opposite parities whenever the Dirac equation is covariant
under $x\rightarrow -x$. 

The charge current density (\ref{currentden}) in this representation takes the form
\begin{equation}\label{j1}
J=c \left(\phi_{+}^{*}\phi_{-} + \phi_{-}^{*}\phi_{+}\right)\,.
\end{equation}

Finally, under the chiral transformations, the spinor becomes
\begin{equation}
\phi_{\chi}= \gamma^{5} \phi =\left( \begin{array}{c} \phi_{-} \\
\phi_{+}
\end{array} \right) \, , \label{eq12b}
\end{equation}
\noindent which effectively interchanges the upper and lower components.

\subsection{The Sturm-Liouville problem}

Using the expression for $\phi_{-}$ obtained from (\ref{eq11b}) with
$E\not =-mc^{2}+\Delta$, \textit{viz}.
\begin{equation}
\phi_{-}= -i\frac{\hbar c \phi^{\prime}_{+}}{E+mc^{2}-\Delta }\label{eq13}
\end{equation}
\noindent and inserting it into Eq. (\ref{eq11a}) one arrives at the
following second-order differential equation for $\phi_{+}$:
\begin{equation}
\begin{split}
-\hbar^{2}c^{2}\phi^{\prime \prime}_{+}&-\hbar c\Delta ^{\prime
}\frac{\hbar c \phi^{\prime}_{+}}{E+mc^{2}-\Delta
}\,\\ & -(E-mc^{2}
-\Sigma)(E+mc^{2}-\Delta )\phi_{+}=0 .\label{psiplus2}
\end{split}
\end{equation}
\noindent The isolated case of Sturm-Liouville problem with $E=-mc^{2}+\Delta$ is derived directly from the original first-order equations (\ref{eq11a}) and (\ref{eq11b}). Here, the isolated solution results in a vanishing upper component and a constant lower component, independently of the functional forms of $\Sigma$ and $\Delta$. Such behavior leads to non-normalizable solutions for bound states, while the resulting zero charge current density confirms the absence of scattering states. 

For $\Delta=0$ with $E\not=-mc^{2}$, Eqs. (\ref{eq13}) and
(\ref{psiplus2}) reduce to
\begin{equation}
\phi_{-}= -i\frac{\hbar c \phi^{\prime}_{+}}{E+mc^{2}
}\, ,\label{eq15a}
\end{equation}
\begin{equation}
-\frac{\hbar^{2}}{2m}\,\phi^{\prime
\prime}_{+}\,+\frac{(E+mc^{2})\Sigma}{2mc^{2}}
\phi_{+}=\frac{E^{2}-m^{2}c^{4}}{2mc^{2}}\,\phi_{+}
\, .\label{eq15b}
\end{equation}
\noindent Under these conditions, substituting Eq.~(\ref{eq15a}) into the current density \eqref{j1} allows $J$ to be expressed solely in terms of the upper component:
\begin{equation}\label{j2}
J=\frac{\hbar c^{2}}{i \left(E+mc^{2}\right)} \left(\phi_{+}^{*}\frac{d\phi_{+}}{dx}-\phi_{+}\frac{d{\phi}_{+}^{*}}{dx}\right)\,.
\end{equation}
\noindent Equation \eqref{j2} is formally equivalent (up to an overall constant factor) to the Schr\"odinger probability current for $\phi_{+}$. Therefore, the reduced equation \eqref{eq15b} and the current \eqref{j2} provide an exact Sturm--Liouville (Schr\"odinger-like) formulation of the relativistic problem for $\Delta=0$. In this framework, the dynamics are described by a single spinor component with an effec\-tive potential uniquely determined by the scalar and vector interactions. 

Although we have focused on the $\Delta=0$ sector, the complementary case $\Sigma=0$ can be obtained directly from the above results by using the chiral transformation discussed in Secs.~\ref{sec:sub:ct} and \ref{sec:sub:eom}. In $1+1$ dimensions this transformation implements $\phi_{\pm}\rightarrow\phi_{\mp}$, $\Sigma\leftrightarrow\Delta$, and $m\rightarrow -m$. Consequently, the Sturm--Liouville reduction for $\Sigma=0$ follows from Eqs.~\eqref{eq15a}--\eqref{j2} by applying these replacements consistently. In particular, at the level of the reduced second-order equation and the associated current, this mapping amounts to exchanging the roles of the upper and lower components and replacing the factor $(E+mc^{2})$ by $(E-mc^{2})$, yielding the analogous Schr\"odinger-like description for $\phi_{-}$ in the $\Sigma=0$ sector.

\subsection{Scattering boundary conditions and definition of transmission probability}
\label{subsec:dirac_scattering_bc}

For scattering states, we assume short-range interactions such that $\Sigma(x) \to 0$ as $x \to \pm\infty$. Consequently, the effective potential vanishes asymptotically, and the solutions approach plane waves. Dividing Eq.~(\ref{eq15b}) by $\hbar^{2}c^{2}$, we recast it into the standard Schrödinger-like form:
\begin{equation}
\left[-\frac{d^{2}}{dx^{2}}+U_{\rm eff}(x;E)\right]\phi_{+}(x)=k^{2}\phi_{+}(x)\,,
\label{eq:schr_form_spin}
\end{equation}
\noindent where the effective energy is identified as
\begin{equation}
k^{2}=\frac{E^{2}-m^{2}c^{4}}{\hbar^{2}c^{2}}\,,
\label{eq:eeff_spin}
\end{equation}
\noindent and the energy-dependent effective potential is given by
\begin{equation}
U_{\rm eff}(x;E)=\frac{E+mc^{2}}{\hbar^{2}c^{2}}\;\Sigma(x)\,.
\label{eq:Ueff_spin}
\end{equation}
\noindent Scattering states correspond to $k \in \mathbb{R}$, which implies $|E| > mc^{2}$. Thus, the Dirac continuum consists of two branches: the positive-energy continuum ($E > mc^{2}$, particle sector) and the negative-energy continuum ($E < -mc^{2}$, antiparticle sector).

Accordingly, a left-incident scattering solution in the positive-energy continuum is defined by the asymptotic behavior
\begin{eqnarray}
\phi_{+}(x) &\sim & e^{ikx}+R(E)\,e^{-ikx},\qquad x\to -\infty, \label{eq:asymp_left0}\\
\phi_{+}(x) &\sim & T(E)\,e^{ikx},\qquad\qquad\ \ \ x\to +\infty, \label{eq:asymp_right0}
\end{eqnarray}
\noindent where $R(E)$ and $T(E)$ are the reflection and transmission amplitudes, respectively. The lower component is subsequently obtained from Eq.~(\ref{eq15a}). 

The transmission and reflection probabilities are defined from the probability current density. Expressing the current solely in terms of $\phi_{+}$ via Eq.~(\ref{j2}), the plane-wave components yield:
\begin{eqnarray}\label{eq:currents_spin}
J_{\rm inc} &=& \frac{\hbar c^{2}k}{E+mc^{2}}\,,\\
J_{\rm ref} &=& -\frac{\hbar c^{2}k}{E+mc^{2}}|R(E)|^{2}\,,\\
J_{\rm tr}  &=& \frac{\hbar c^{2}k}{E+mc^{2}}|T(E)|^{2}\,.
\end{eqnarray}
\noindent The sign of the current identifies the incident, reflected, and transmitted fluxes in the standard manner. While the same formal expressions apply to the negative-energy continuum ($E<-mc^{2}$), the interpretation of flux directions must account for the sign change of $E+mc^{2}$; this sector is treated analogously and will not be detailed here. 

The transmission and reflection probabilities are then simply
\begin{eqnarray}\label{eq:TR_spin}
\mathcal{T}(E) &=& \frac{J_{\rm tr}}{J_{\rm inc}}=|T(E)|^{2}\,,\\
\mathcal{R}(E) &=& \frac{|J_{\rm ref}|}{J_{\rm inc}}=|R(E)|^{2}\,,
\end{eqnarray}
\noindent satisfying the unitary condition $\mathcal{R}(E)+\mathcal{T}(E)=1$. 

Finally, discrete and unstable spectra can be extracted via the analytic continuation of $T(E)$. Bound states correspond to purely imaginary, $k=i\vert\kappa\vert$, yielding square-integrable solutions. Conversely, resonances (quasi-normal modes) are associated with poles in the lower-half complex $k$-plane, characterized by purely outgoing boundary conditions at both spatial infinities.
 
\section{Review of SUSYQM}
\label{sec:revSUSYQM}

\subsection{Supersymmetric factorization and partner Hamiltonians}

A central idea in supersymmetric quantum mechanics (SUSYQM) is that a broad class of
one-dimensional Schr\"odinger Hamiltonians can be factorized into first-order differential operators.
Working in rescaled units (or after the standard shift that sets the lowest eigenvalue to zero),
consider a Hamiltonian $H_1$ depending on a set of parameters $a_1$,
\begin{equation}\label{eq:H1}
H_1\,\Phi_n^{(1)}(x;a_1)=\mathcal{E}_n^{(1)}\,\Phi_n^{(1)}(x;a_1)\,,
\end{equation}
\noindent where
\begin{equation}
H_{1}=-\frac{d^2}{dx^2}+V_1(x;a_1)\,.
\end{equation}
\noindent This spectral problem admits the factorization
\begin{equation}\label{eq:factorization}
H_1=\mathcal{A}^\dagger(a_1)\,\mathcal{A}(a_1),
\end{equation}
\noindent where the SUSY operators are defined in terms of a superpotential $W(x;a_1)$ as
\begin{eqnarray}\label{eq:Aoperators}
\mathcal{A}^\dagger(a_1) &=& -\frac{d}{dx}+W(x;a_1)\,,\\
\mathcal{A}(a_1) &=& \frac{d}{dx}+W(x;a_1)\,.
\end{eqnarray}
\noindent The potential $V_1$ is then expressed directly in terms of $W$:
\begin{equation}\label{eq:V1}
V_1(x;a_1)=W^2(x;a_1)-\frac{dW(x;a_1)}{dx}\,.
\end{equation}
\noindent The superpotential can be constructed from the ground-state wavefunction (assumed nodeless) as
\begin{equation}\label{eq:Wfromground}
W(x;a_1)=-\frac{d}{dx}\ln \Phi_0^{(1)}(x;a_1)\,.
\end{equation}

The supersymmetric partner Hamiltonian is obtained by reversing the operator order,
\begin{equation}\label{eq:H2}
H_2=\mathcal{A}(a_1)\,\mathcal{A}^\dagger(a_1)=-\frac{d^2}{dx^2}+V_2(x;a_1)\,,
\end{equation}
\noindent with partner potential
\begin{equation}\label{eq:V2}
V_2(x;a_1)=W^2(x;a_1)+\frac{dW(x;a_1)}{dx}\,.
\end{equation}
\noindent The spectra of $H_1$ and $H_2$ are identical up to the possible absence of the ground state in one sector.
In particular, with the conventional SUSY shift $\mathcal{E}_0^{(1)}=0$,
\begin{equation}\label{eq:spectraldegeneracy}
\mathcal{E}_0^{(1)}=0,\qquad \mathcal{E}_n^{(2)}=\mathcal{E}_{n+1}^{(1)}\quad (n=0,1,2,\dots)\,,
\end{equation}
\noindent and eigenfunctions at the same energy are related through $\mathcal{A}^\dagger$,
\begin{equation}\label{eq:eigenfunctionrelation}
\Phi_{n+1}^{(1)}(x;a_1)=\left(\mathcal{E}_n^{(2)}\right)^{-1/2}\mathcal{A}^\dagger(a_1)\,\Phi_{n}^{(2)}(x;a_1)\,.
\end{equation}

\subsection{Shape invariance}

The factorization procedure can be iterated, generating a hierarchy of partner potentials.
A particularly important solvable class is obtained when the partner potentials are \emph{shape invariant},
i.e., when $V_2$ has the same functional form as $V_1$ but with transformed parameters, up to an additive constant:
\begin{equation}\label{eq:shapeinvariance}
V_2(x;a_1)=V_1(x;a_2)+C(a_1)\,,
\end{equation}
\noindent where $a_2=f(a_1)$ for some parameter map $f$, and $C(a_1)$ is independent of $x$.
In this case, many spectral and scattering properties can be obtained algebraically via the parameter sequence
$a_{n+1}=f(a_n)$, a fact that will be exploited below in the context of scattering.

\subsection{Scattering states and SUSY relations for $T$ and $R$}
\label{sec:sbc}

We now specialize to one-dimensional scattering from a potential with continuous spectrum ($\mathcal{E}=k^2>0$ in the rescaled convention).
Let $\Phi^{(2)}(k,x)$ be a scattering eigenfunction of $H_2$ with asymptotic plane-wave behavior
\begin{align}
\Phi^{(2)}(k,x\to -\infty)&\sim e^{ikx}+R_2(k)\,e^{-ikx}\,, \label{eq:asymp2minus}\\
\Phi^{(2)}(k,x\to +\infty)&\sim T_2(k)\,e^{ikx}\,. \label{eq:asymp2plus}
\end{align}
\noindent Equations (\ref{eq:asymp2minus}) and (\ref{eq:asymp2plus}) define the left-incident scattering problem for the partner
Hamiltonian $H_2$, with $T_2(k)$ and $R_2(k)$ the transmission and reflection amplitudes.
The SUSY intertwining then maps these solutions into scattering states of $H_1$, implying simple
algebraic relations between the corresponding amplitudes (and therefore between the associated transmission and reflection probabilities).
Using the intertwining relation between partner wavefunctions (the scattering analogue of
Eq.~\eqref{eq:eigenfunctionrelation}), one can determine the asymptotics of the corresponding $H_1$
scattering state $\Phi^{(1)}(k,x)\propto \mathcal{A}^\dagger(a_1)\Phi^{(2)}(k,x)$. Denoting the asymptotic limits
of the superpotential by
\begin{equation}\label{eq:Wasymp}
W_\pm(a_1)=\lim_{x\to\pm\infty}W(x;a_1)\,,
\end{equation}
\noindent one finds the algebraic relations between the transmission and reflection amplitudes:
\begin{align}
T_1(k)&=\frac{W_+(a_1)-ik}{W_-(a_1)-ik}\,T_2(k)\,, \label{eq:Trel}\\
R_1(k)&=\frac{W_-(a_1)+ik}{W_-(a_1)-ik}\,R_2(k)\,. \label{eq:Rrel}
\end{align}
\noindent (For simplicity, one often assumes asymptotically symmetric situations such that the relevant limits are consistent,
e.g. $W_+^2=W_-^2$, which is sufficient for many solvable wells.)

If, in addition, the potential is shape invariant, then the partner problem at parameters $a_1$ is equivalent
(up to an additive constant) to the original functional form at parameters $a_2=f(a_1)$. Consequently,
Eqs.~\eqref{eq:Trel}--\eqref{eq:Rrel} become recursion relations entirely in terms of the same functional superpotential
with shifted parameters:
\begin{align}
T(k;a_1)&=\frac{W_+(a_1)-ik}{W_-(a_1)-ik}\,T(k;a_2)\,, \label{eq:Trec}\\
R(k;a_1)&=\frac{W_-(a_1)+ik}{W_-(a_1)-ik}\,R(k;a_2)\,. \label{eq:Rrec}
\end{align}
\noindent These recursion relations can be iterated along the parameter chain $a_{n+1}=f(a_n)$ until a value
$a_N$ is reached for which the scattering data are known (e.g. a trivial or free limit).
Iterating from $a_1$ to a parameter value $a_N$ yields closed product expressions:
\begin{equation}\label{eq:Tproduct}
T(k;a_1)=\prod_{n=1}^{N-1}\frac{W_+(a_n)-ik}{W_-(a_n)-ik}\;T(k;a_N)\,,
\end{equation}
\begin{equation}\label{eq:Rproduct}
R(k;a_1)=\prod_{n=1}^{N-1}\frac{W_-(a_n)+ik}{W_-(a_n)-ik}\;R(k;a_N)\,.
\end{equation}
\noindent These formulas provide an algebraic route to the scattering amplitudes for a large class
of exactly solvable, shape-invariant potentials.

\section{Application: P\"oschl--Teller scalar and vector interactions}
\label{sec:PT_application}

\subsection{Effective P\"oschl--Teller potential in the symmetry limits}\label{sec:sub:ptp}
Let us consider
\begin{equation}\label{eq:ptp}
 \Sigma=-2V_{0}\,\sech^{2}(\omega x)\,.
\end{equation}
\noindent Substituting (\ref{eq:ptp}) into (\ref{eq:schr_form_spin}) the resulting equation shows that, in the symmetry limit considered here, the Dirac problem reduces to
a Schr\"odinger-like equation with an energy-dependent P\"oschl--Teller effective interaction
\begin{equation}\label{eq:pt1}
U_{\rm eff}(x;E)= -\frac{2(E+mc^{2})}{\hbar^{2}c^{2}}\;V_{0}\,\sech^{2}(\omega x)\,.
\end{equation}
\noindent Since $\sech^{2}(\omega x)\ge 0$, the qualitative nature of the effective potential is controlled by the sign of the product $(E+mc^{2})V_{0}$.

For $V_{0}>0$ and $E>-mc^{2}$ the effective potential is attractive,
$U_{\rm eff}(x;E)<0$, and has the shape of a localized well. In this case one can study scattering states in the positive-energy continuum
($E>mc^{2}$) as well as a finite set of bound states (with $|E|<mc^{2}$), which appear as square-integrable solutions of the
Schr\"odinger-like equation and can be extracted from the pole structure of the analytically continued
transmission amplitude. For $V_{0}<0$ and $E>-mc^{2}$ the effective potential becomes repulsive,
$U_{\rm eff}(x;E)>0$, corresponding to a localized barrier. Scattering still takes place for continuum energies,
and the barrier configuration may support
resonant (quasi-bound) states, characterized by complex poles of the transmission amplitude in the lower-half
complex momentum plane. In many wave settings, these poles are also referred to as quasi-normal modes, as they correspond to purely outgoing conditions at both spatial infinities. 

In contrast, for the negative-energy continuum $E<-mc^{2}$ the above classification is
reversed: $V_{0}>0$ yields a barrier while $V_{0}<0$ yields a well.

In all cases, once the effective interaction is cast in the P\"oschl--Teller form, the scattering problem can
be treated analytically with the SUSYQM/shape-invariance machinery summarized in Sec.~\ref{sec:revSUSYQM}, yielding closed
expressions for the transmission amplitude and allowing one to extract bound and resonant information from its
poles.

\subsection{SUSYQM construction and transmission amplitude}
\label{sec:sub:ta}

To compute the transmission amplitude associated with the effective interaction \eqref{eq:pt1}, we now
make use of the exactly solvable SUSYQM construction for the P\"oschl--Teller family. In particular, the
effective well/barrier proportional to $\sech^{2}(\omega x)$ can be generated from the superpotential
\begin{equation}\label{eq:W_tanh}
W(x)=A\,\tanh(\omega x)\,,
\end{equation}
\noindent where $A$ controls the strength and $\omega>0$ sets the length scale. Defining $H_-=\mathcal{A}^\dagger\mathcal{A}$ and $H_+=\mathcal{A}\mathcal{A}^\dagger$ and using $W'(x)=A\omega\,\sech^2(\omega x)$ and $\tanh^2(\omega x)=1-\sech^2(\omega x)$, the partner potentials
$V_{\mp}(x)=W^2(x)\mp W'(x)$ read
\begin{equation}
V_{\mp}=A^{2}-A(A\pm\omega)\,\sech^2(\omega x)\,.
\end{equation}
\noindent Since $V_\pm(x)\to A^2$ as $x\to\pm\infty$, it is convenient to subtract this constant asymptote and work with shifted Hamiltonians
\begin{equation}\label{eq:Hshift}
\widetilde H_\pm \equiv H_\pm - A^2 = -\frac{d^2}{dx^2}+U_\pm(x)\,,
\end{equation}
\noindent with
\begin{equation}
U_\pm(x)\equiv V_\pm(x)-A^2\,,
\end{equation}
\noindent so that $U_\pm(x)\to0$ at infinity. Explicitly,
\begin{equation}\label{eq:UminusUplus}
U_{\mp}(x) = -A(A\pm\omega)\,\sech^2(\omega x)\,.
\end{equation}
\noindent Introducing the dimensionless parameter $\lambda=A/\omega$, one recognizes the attractive P\"oschl--Teller family,
\begin{equation}\label{eq:PTlambda1}
U_{\mp}(x) = -\lambda(\lambda\pm 1)\omega^2\,\sech^2(\omega x)\,,
\end{equation}
\noindent which is shape invariant in the form
\begin{equation}\label{eq:shape_inv_tanh}
U_+(x;A)=U_-(x;A-\omega).
\end{equation}
\noindent In the Dirac reduction of Sec.~\ref{sec:sub:ptp}, the effective interaction (\ref{eq:pt1}) is of the same functional form
as $U_-(x)$, and the two can be identified by matching their depths (which introduces an energy dependence in
the effective coupling).

Let $\psi_\pm(k,x)$ be scattering solutions of the shifted problems
\begin{equation}\label{eq:scatt_eq_shift}
\left[-\frac{d^2}{dx^2}+U_\pm(x)\right]\psi_\pm(k,x)=k^2\psi_\pm(k,x)\,,
\end{equation}
\noindent with $k^2=\mathcal{E}-A^2$ ($k\in\mathbb{R}^{+}$), and normalized by the asymptotic conditions
\begin{align}
\psi_\pm(k,x) &\sim e^{ikx}+R_\pm(k)e^{-ikx},\qquad x\to-\infty\,, \label{eq:asymp_left}\\
\psi_\pm(k,x) &\sim T_\pm(k)e^{ikx},\qquad\qquad\ \ x\to+\infty\,. \label{eq:asymp_right}
\end{align}
\noindent The partner scattering states are intertwined (up to an overall constant) by
\begin{equation}\label{eq:intertwine}
\psi_-(k,x)\propto \mathcal{A}^\dagger\,\psi_+(k,x)\,.
\end{equation}
\noindent For $W(x)=A\tanh(\omega x)$, the superpotential approaches constants at spatial infinity,
\begin{equation}\label{eq:W_asymp}
W(\pm\infty)=\pm A\,.
\end{equation}
\noindent By matching Eqs.~\eqref{eq:asymp_left}--\eqref{eq:asymp_right}, one obtains
\begin{align}
T_-(k) &= \frac{-ik+W(+\infty)}{-ik+W(-\infty)}\,T_+(k)
       = \frac{A-ik}{-A-ik}\,T_+(k), \label{eq:T_relation}\\
R_-(k) &= \frac{ik+W(-\infty)}{-ik+W(-\infty)}\,R_+(k)
       = \frac{ik-A}{-ik-A}\,R_+(k). \label{eq:R_relation}
\end{align}
\noindent Because $|A-ik|=|A+ik|$, the prefactors have unit modulus. Therefore SUSY changes only phases and preserves the transmission and reflection probabilities:
\begin{equation}
|T_-(k)|^2=|T_+(k)|^2,\qquad |R_-(k)|^2=|R_+(k)|^2.
\end{equation}

Using shape invariance \eqref{eq:shape_inv_tanh}, one may identify $T_+(k;A)=T_-(k;A-\omega)$ and rewrite \eqref{eq:T_relation} as a closed recursion:
\begin{equation}\label{eq:T_recursion}
T(k;A)=\frac{A-ik}{-A-ik}\,T\!\left(k;A-\omega\right)\,.
\end{equation}
\noindent Iterating $N$ times with $A=N\omega$ reduces the problem to the free case $A\to0$, for which $T(k;0)=1$ and $R(k;0)=0$. Hence
\begin{eqnarray}\label{eq:T_product}
T\left(k;N\omega\right) &=& \prod_{n=1}^{N}\frac{k+in\omega}{k-in\omega}\,,\\
R\left(k;N\omega\right) &=& 0\,.
\end{eqnarray}
\noindent Therefore, for integer $\lambda=N$ the P\"oschl--Teller well generated by $W=A\tanh(\omega x)$ is reflectionless and
$\mathcal{T}(k)\equiv|T(k)|^2=1$.

For completeness, for generic $\lambda\in\mathbb{R}$ and $q=k/\omega$, the transmission amplitude for
$U_-(x)$ can be expressed as (see \ref{appendix:1:1} for further details)
\begin{equation}\label{eq:T_gamma}
T(k)=
\frac{\Gamma\left(\lambda+1-iq\right)\,
      \Gamma\left(-\lambda-iq\right)}
     {\Gamma\left(1-iq\right)\,
      \Gamma\left(-iq\right)}\,.
\end{equation}
\noindent The transmission probability $\mathcal{T}(k)$, reduces to (see \ref{appendix:1:2} for further details):
\begin{equation}\label{eq:T_sinh_sin}
\mathcal{T}(k)=\frac{p^2}{1+p^2}\,,
\end{equation}
\noindent where $p=\frac{\sinh\pi q}{\sin\pi \lambda}$. Notably, the system becomes perfectly transparent ($\mathcal{T}=1$) whenever $\lambda$ is an integer.

\subsection{Bound states from the poles of the transmission amplitude}
\label{subsec:well_bound_state}

For $V_{0}>0$ the effective interaction \eqref{eq:pt1} is attractive and supports a finite number of
bound states. In the present approach these states can be obtained directly from the pole structure of the
transmission amplitude. To make contact with the P\"oschl--Teller parametrization introduced in Sec.~\ref{sec:sub:ta}, we
compare Eq.~(\ref{eq:pt1}) with the standard attractive P\"oschl--Teller well (\ref{eq:PTlambda1}), which implies the identification
\begin{equation}\label{eq:lambda_match}
\lambda(\lambda+1)\omega^{2}
=\frac{2(E+mc^{2})}{\hbar^{2}c^{2}}\,V_{0}.
\end{equation}
\noindent Solving Eq.~\eqref{eq:lambda_match} for $\lambda$ gives
\begin{equation}\label{eq:lambda_explicit}
\lambda(E)=\frac{1}{2}\left( -1\pm\sqrt{1+\frac{8(E+mc^{2})V_{0}}{\hbar^{2}c^{2}\omega^{2}}} \right)\,.
\end{equation}
\noindent For the attractive case $V_{0}>0$ (with $E>-mc^{2}$), we take the ``$+$'' branch in
Eq.~\eqref{eq:lambda_explicit} so that $\lambda\ge 0$. Notice that, because the Dirac reduction produces an energy-dependent effective potential, the parameter $\lambda$ is not a fixed coupling: it is determined self-consistently by the Dirac energy $E$ through
Eqs.~\eqref{eq:lambda_match}--\eqref{eq:lambda_explicit}. 

\begin{figure}[ht]
\centering
\begin{subfigure}{0.75\linewidth}
\centering
\includegraphics[width=\linewidth]{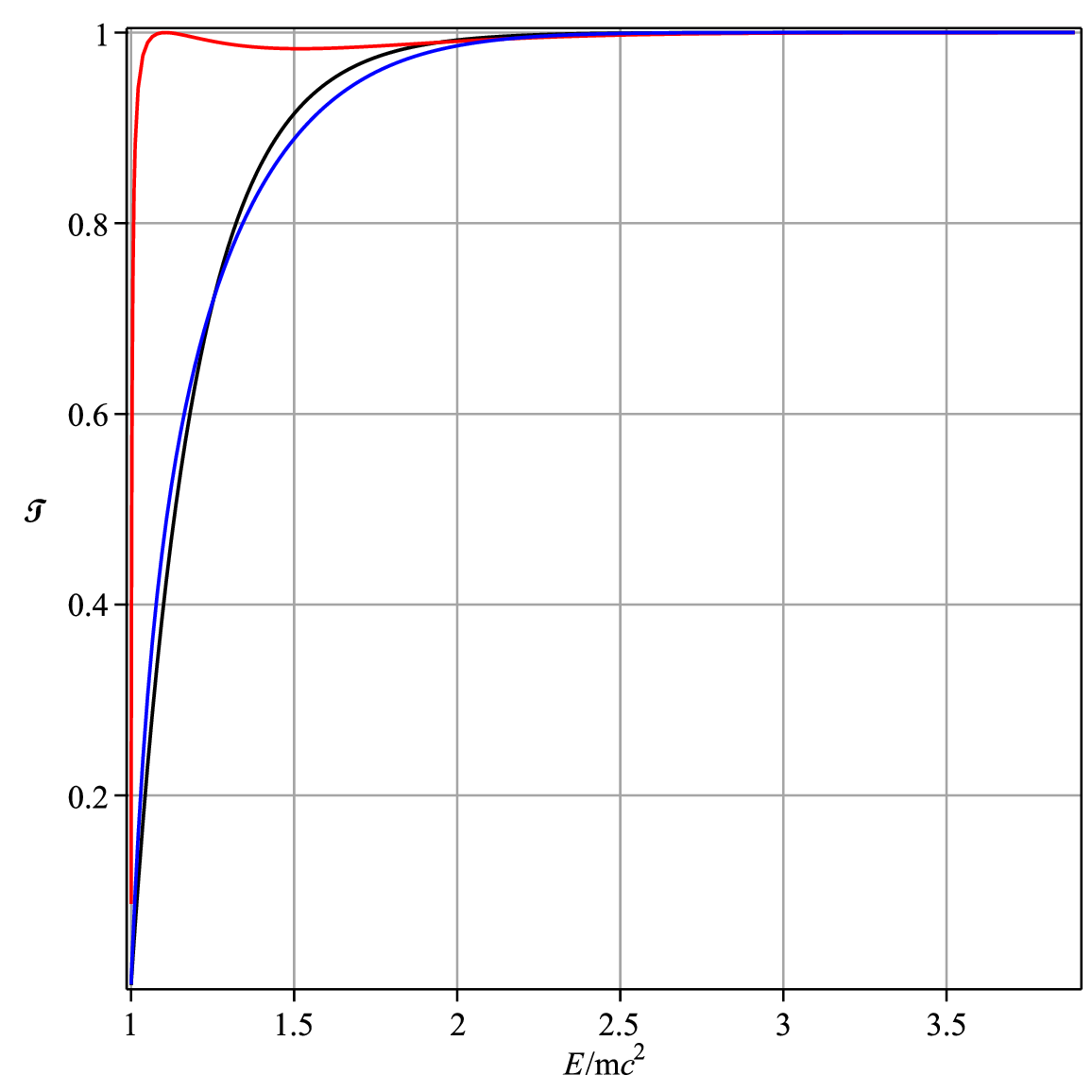}
\caption{}
\label{fig:TRa}
\end{subfigure}
\begin{subfigure}{0.75\linewidth}
\centering
\includegraphics[width=\linewidth]{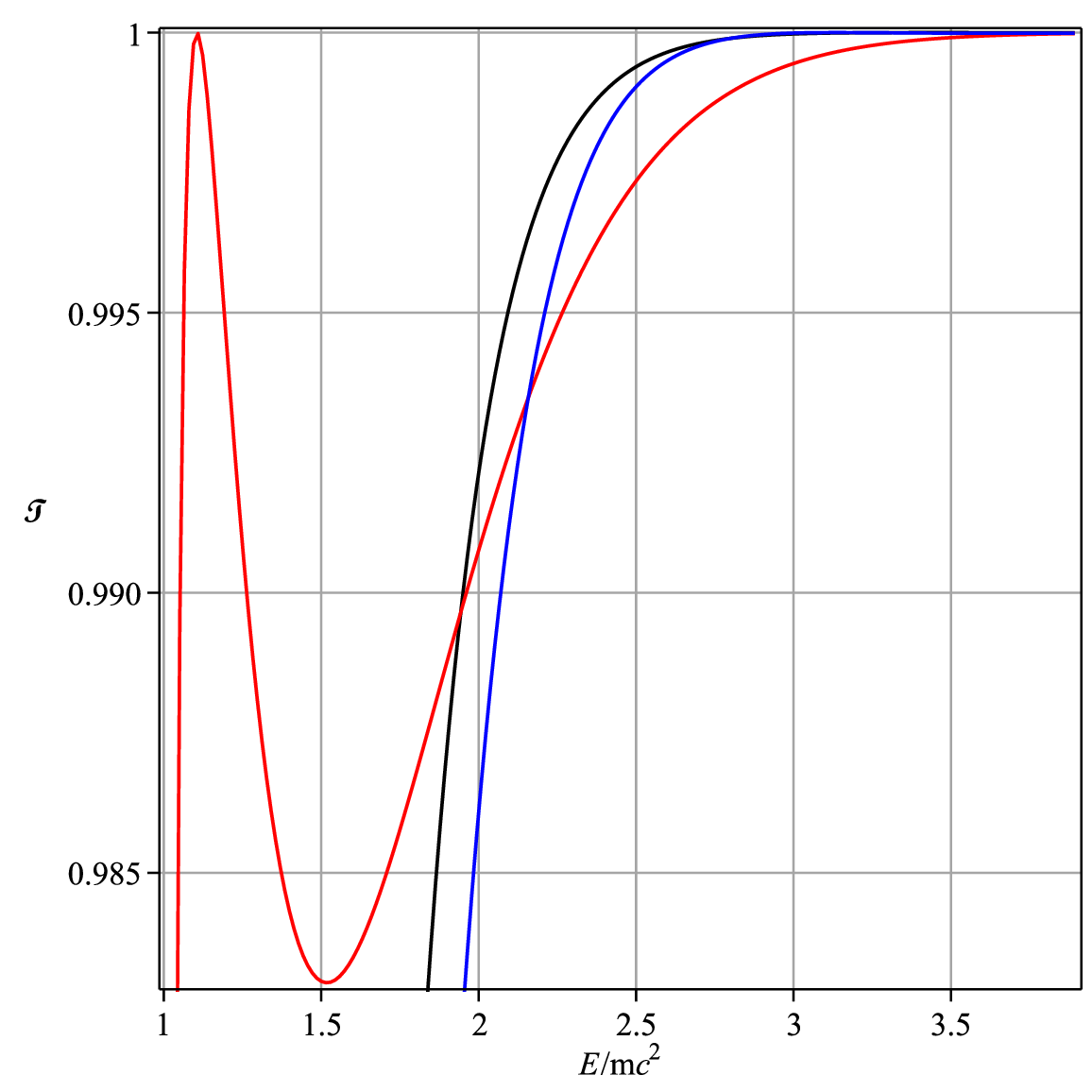}
\caption{}
\label{fig:TRb}
\end{subfigure}
\caption{\label{fig:TR} Transmission probability $\mathcal{T}$ for the Dirac-matched P\"oschl--Teller effective well with
$\omega=2$ and $V_{0}=0.9$ (black), $V_{0}=1.9$ (red), and $V_{0}=2.9$ (blue) (in units where $\hbar=c=m=1$). (a) Full energy range. (b) Magnified view of the high-transmission region $\mathcal{T}\in[0.983,1.0]$.}
\end{figure}

\begin{figure}[t]
\centering
\includegraphics[width=0.35\textwidth]{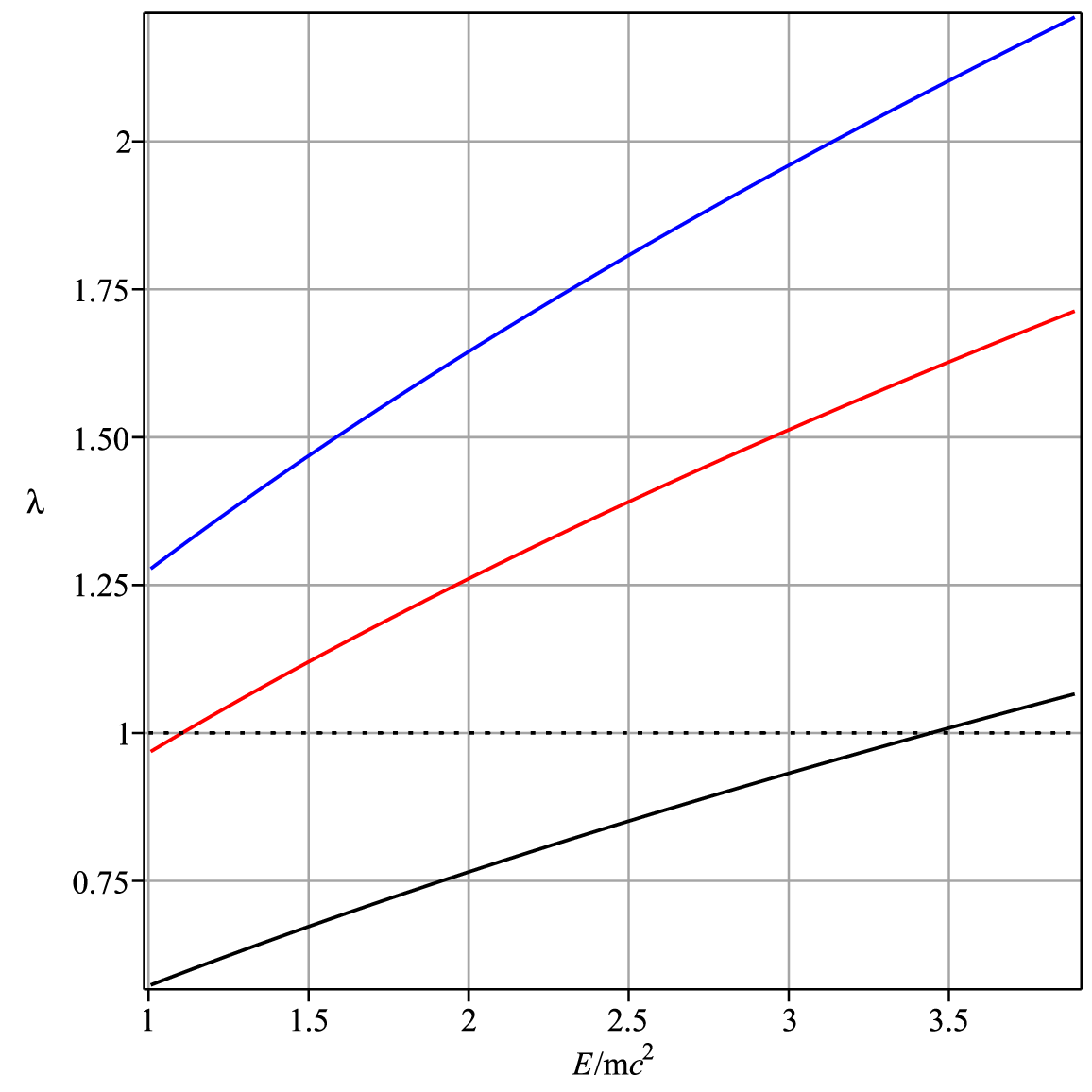}
\caption{\label{fig:lambda} Energy dependence of the P\"oschl--Teller parameter $\lambda(E)$ from the Dirac matching condition
for $\omega=2$ and $V_{0}=0.9$ (black), $V_{0}=1.9$ (red), and $V_{0}=2.9$ (blue) (in units where $\hbar=c=m=1$).}
\end{figure}

Before turning to the pole analysis, it is instructive to illustrate how the Dirac matching $\lambda=\lambda(E)$ affects the transmission probability in the continuum. Figure~\ref{fig:TR} shows the transmission probability for the Dirac-matched P\"oschl--Teller well (in units $\hbar=c=m=1$) for $\omega=2$ and $V_{0}=0.9,\,1.9,\,2.9$. Close to the positive-energy threshold $E\gtrsim mc^{2}$ ($k\to 0$), $\mathcal{T}(E)$ is most sensitive to the effective coupling and may differ substantially among the curves. As $E$ increases, $k$ (and therefore $q=k/\omega$) grows and the hyperbolic factor $\sinh(\pi q)$ rapidly dominates, driving $\mathcal{T}(E)\to 1$ for all values of $V_{0}$, as expected for short-range interactions. The non-monotonic dependence on $V_{0}$ observed near threshold is clarified by Fig.~\ref{fig:lambda}, which displays the energy-dependent parameter $\lambda(E)$ obtained from the Dirac matching. In the closed-form P\"oschl--Teller expression, the departure of $\mathcal{T}(E)$ from unity is governed by the factor $\sin^{2}\!\bigl[\pi\lambda(E)\bigr]$; when $\lambda(E)$ passes close to an integer, $\sin^{2}\!\bigl[\pi\lambda(E)\bigr]\approx 0$ and the transmission becomes almost perfect. This explains why, for the present parameters, the case $V_{0}=1.9$ exhibits enhanced transmission near threshold because $\lambda(E)$ lies close to $\lambda\simeq 1$ in that energy range, whereas for $V_{0}=0.9$ and $V_{0}=2.9$ the corresponding $\lambda(E)$ stays farther from integer values and the low-energy transmission is smaller.

We now proceed to extract the discrete spectrum from the pole structure of the analytically continued
transmission amplitude. Bound states correspond to square-integrable solutions of \eqref{eq:schr_form_spin} and are obtained by
continuing the scattering problem to imaginary wavenumber $k=i\vert\kappa\vert$, so that the asymptotic behavior becomes exponentially decaying at $x\to\pm\infty$.
In the P\"oschl--Teller case, the transmission amplitude can be written in terms of gamma functions [Eq.~(\ref{eq:T_gamma})]. Since $\Gamma(z)$ has simple poles at $z=0,-1,-2,\dots$, poles of $T(k)$ arise whenever
\begin{equation}\label{eq:pole_condition}
-\lambda-iq=-n,\qquad n=0,1,2,\dots,
\end{equation}
\noindent which yields
\begin{equation}\label{eq:k_poles}
k_{n}=i\omega(\lambda-n)\,.
\end{equation}
\noindent The bound-state condition $\kappa>0$ therefore requires $n<\lambda$, implying a finite number of levels.

Finally, the bound-state energies of the original Dirac problem follow from combining the pole condition with
the dispersion relation. In particular, with $k=i\vert\kappa\vert$ and \eqref{eq:k_poles}, one obtains
\begin{equation}\label{eq:bound_energy_condition}
\sqrt{m^{2}c^{4}-E^{2}}=\hbar c\omega\,a_{n}\,,
\end{equation}
\noindent with $a_{n}=\lambda-n>0$. Solving these relations yields the bound-state
spectrum and reproduces the known results for the Dirac--P\"oschl--Teller system \cite{EPL77:20009:2007}. In this way, scattering and
bound states are described within a unified framework based on the analytic properties of $T(k)$.

\subsection{Barrier configurations: resonances and quasi-normal modes}
\label{subsec:barrier_qnm}

In the barrier regime, $V_{0}<0$, the effective interaction in Eq.~\eqref{eq:pt1} becomes repulsive,
$U_{\rm eff}(x;E)>0$, and therefore has the form of a localized P\"oschl--Teller ba\-rrier.
In this case the discrete spectrum is not formed by square-integrable bound states.
Instead, one may look for \emph{unstable} states associated with poles of the analytically
continued transmission amplitude in the complex $k$-plane. In standard scattering
terminology, these poles are interpreted as resonances; in wave problems they are also
referred to as quasi-normal modes when they are selected by purely outgoing conditions
at both spatial infinities \cite{PRD30:295:1984}.

As in Sec.~\ref{subsec:well_bound_state}, we match the effective strength to the P\"oschl--Teller parameter via
Eq.~\eqref{eq:lambda_match}, but now $V_{0}=-|V_{0}|<0$. Solving explicitly for $\lambda$ yields
\begin{equation}\label{eq:lambda_barrier_explicit}
\lambda(E)=\frac{1}{2}\left( -1\pm\sqrt{\,1-\dfrac{8(E+mc^{2})|V_{0}|}{\hbar^{2}c^{2}\omega^{2}}} \right)\,.
\end{equation}
\noindent Therefore, the qualitative pole structure depends on the sign of the discriminant
\begin{equation}\label{eq:disc_barrier}
D(E)\equiv 1-\frac{8(E+mc^{2})|V_{0}|}{\hbar^{2}c^{2}\omega^{2}}\,.
\end{equation}

If $D(E)>0$, then $\lambda(E)$ is real and lies in the interval $-1<\lambda<0$ (for $E>-mc^{2}$),
corresponding to a ``low'' barrier. In this regime the P\"oschl--Teller barrier does not
exhibit resonance poles; instead the analytic continuation leads to poles on the imaginary
axis (antibound/virtual states), similarly to the well case \cite{PLA380:1600:2016}.

If $D(E)<0$, the square root in Eq.~\eqref{eq:lambda_barrier_explicit} becomes imaginary and
$\lambda$ acquires a nonzero imaginary part. It is then convenient to parameterize
\begin{equation}\label{eq:lambda_eta}
\lambda(E)=-\frac{1}{2}\pm i\,\eta(E)\,,
\end{equation}
\noindent with 
\begin{equation}\label{eq:eta}
\eta(E)=\frac{1}{2}\sqrt{\frac{8(E+mc^{2})|V_{0}|}{\hbar^{2}c^{2}\omega^{2}}-1}\,,
\end{equation}
\noindent which corresponds to a ``high'' barrier. In this case the poles of the transmission amplitude
move off the imaginary axis and form two symmetric families in the lower-half complex
momentum plane \cite{PLA380:1600:2016}.

More explicitly, using the gamma-function representation of the transmission amplitude
(analytic continuation of Eq.~(\ref{eq:T_gamma})), the pole conditions arise from the poles of the gamma
functions and can be written as
\begin{equation}\label{eq:qnm_poles_q}
q_{n}^{(\pm)}=\pm \eta(E)-i\left(n+\frac{1}{2}\right), \qquad n=0,1,2,\ldots,
\end{equation}
\noindent where $q=k/\omega$. Hence
\begin{equation}\label{eq:qnm_poles_k}
k_{n}^{(\pm)}=\omega\,q_{n}^{(\pm)}=\pm \omega\,\eta(E)-i\omega\left(n+\frac{1}{2}\right).
\end{equation}
\noindent These poles correspond to complex energies through the dispersion relation
$k^{2}=(E^{2}-m^{2}c^{4})/(\hbar^{2}c^{2})$. Because the effective coupling
(and thus $\eta$) depends on $E$, determining the resonance/QNM frequencies in the Dirac
problem requires solving Eqs.~\eqref{eq:lambda_eta}--\eqref{eq:qnm_poles_k} self-consistently
in the complex plane. This provides a natural starting point for a dedicated QNM analysis,
in the spirit of the inverted-potential viewpoint used in related contexts \cite{PRD30:295:1984}.

\begin{figure}[t]
\centering
\includegraphics[width=0.35\textwidth]{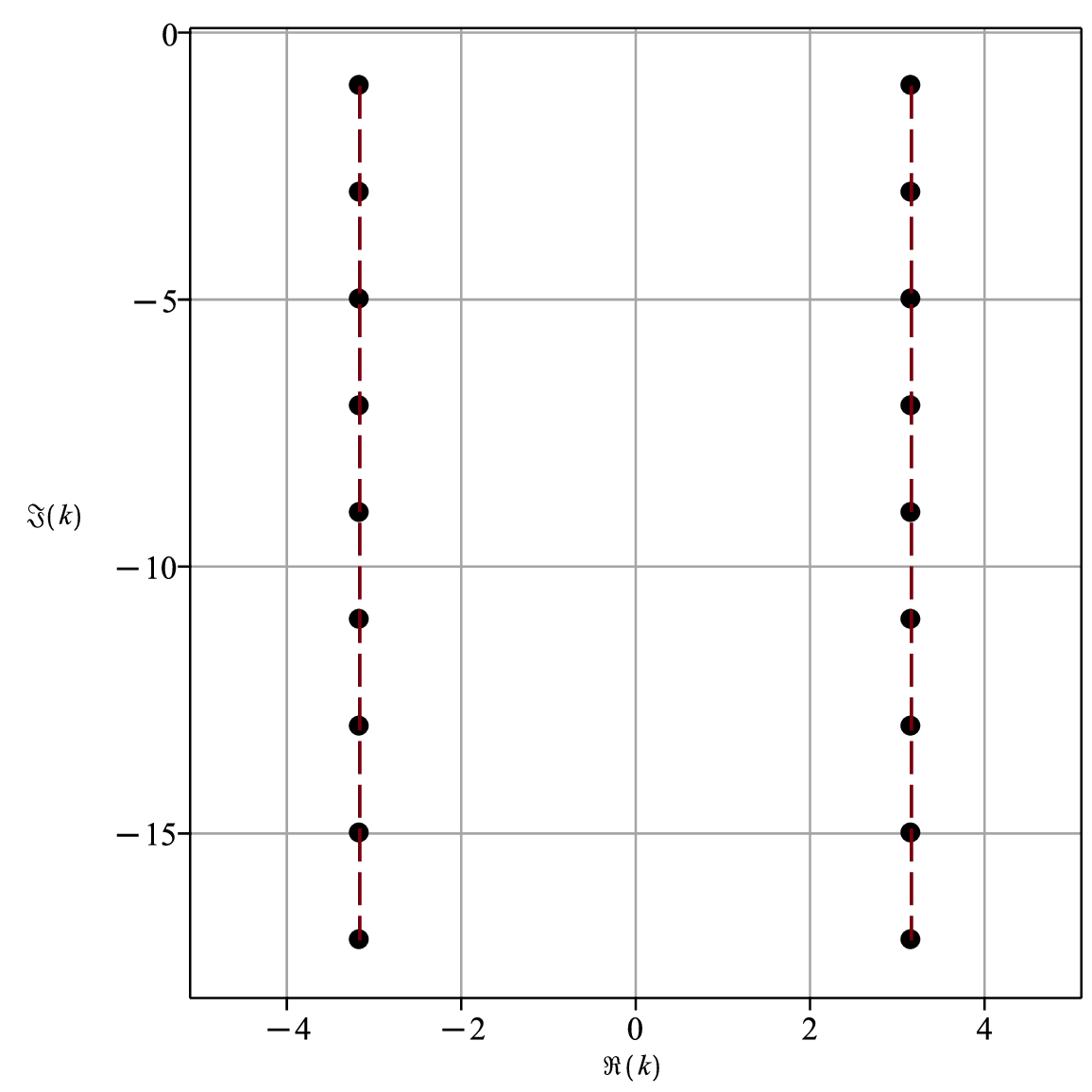}
\caption{\label{fig:poles} Pole pattern in the complex $k$-plane for the Dirac-matched P\"oschl--Teller barrier with
$\omega=2$, $V_{0}=-2.5$, and $E_{0}=1.2$ (in units where $\hbar=c=m=1$), yielding $\eta(E_{0})\simeq 1.58$ and
$k_{n}^{(\pm)}=\pm \omega\eta(E_{0})-i\omega(n+\tfrac12)$ ($n=0,\ldots,8$).}
\end{figure}

The pole pattern displayed in Fig.~\ref{fig:poles} is a direct signature of the high-barrier regime, in which
the discri\-minant $D(E)$ is negative and the P\"oschl--Teller para\-meter becomes complex, $\lambda=-\tfrac12\pm i\eta$ with
$\eta\in\mathbb{R}$. In this case the poles form two symmetric families $k_{n}^{(\pm)}$ in the lower-half
complex $k$-plane: they are located at fixed real parts $\Re(k)=\pm \omega\eta$ and are equally spaced along
the imaginary direction, $\Im(k)=-\omega(n+\tfrac12)$. The symmetry $k\to -k$ reflects the left--right symmetry
of the barrier, while the uniform spacing in $\Im(k)$ implies an increasing damping rate for higher-order
modes. The single parameter $\eta$ controls the horizontal displacement of the pole towers: varying $|V_{0}|$
(or, equivalently, chan\-ging the reference energy $E_{0}$ used in the matching) does not alter the qualitative
structure of the spectrum, but only shifts the two towers away from or towards the imaginary axis.

The boundary between high- and low-barrier beha\-vior is set by $D(E)=0$, i.e.
\begin{equation}
|V_{0}|=\frac{\hbar^{2}c^{2}\omega^{2}}{8(E+mc^{2})}.
\end{equation}
As $D(E)\to 0^{-}$ one has $\eta\to 0$ and the two pole towers approach the imaginary axis, whereas for
$D(E)>0$ the parameter $\eta$ becomes purely imaginary and the poles collapse to $\Re(k)=0$, corresponding to antibound (virtual) states rather than resonances/quasi-normal modes.

\subsection{Pseudospin-symmetry sector ($\Sigma=0$) via the chiral transformation}
\label{subsec:pseudospin_from_sym}

The pseudospin-symmetry limit $\Sigma=0$ can be discussed without repeating the full derivation. As reviewed in Secs.~\ref{sec:sub:ct} and \ref{sec:sub:eom}, the chiral transformation implements the simultaneous mappings $\Delta\rightarrow\Sigma$, $\Sigma\rightarrow\Delta$, $m\rightarrow-m$ and $\phi_{\pm}\rightarrow\phi_{\mp}$. Under this transformation, results obtained in one symmetry sector can be translated to the other by applying these replacements consistently.

Accordingly, once the effective interaction is cast into a P\"oschl--Teller form in the pseudospin sector, the
SUSYQM scattering results obtained in Secs.~\ref{sec:sub:ta}--\ref{subsec:barrier_qnm} can be reused with minimal modifications. For example,
choosing $\Delta(x)=-2V_{0}\sech^{2}(\omega x)$ with $\Sigma=0$ and $E\neq mc^{2}$, Eq.~(\ref{eq:Ueff_spin}) leads to an effective potential
\begin{equation}
U_{\rm eff}^{(\Sigma=0)}(x;E)
=-\frac{2(E-mc^{2})}{\hbar^{2}c^{2}}\,V_{0}\,\sech^{2}(\omega x),
\end{equation}
which differs from the spin-symmetric case only by the replacement $(E+mc^{2})\to(E-mc^{2})$. Consequently, the
P\"oschl--Teller parameter is fixed by the matching condition
\begin{equation}
\lambda(\lambda+1)\omega^{2}=\frac{2(E-mc^{2})}{\hbar^{2}c^{2}}\,V_{0},
\end{equation}
and the transmission amplitude, its analytic continuation, and the extraction of bound states and
res\-o\-nance/quasi-normal modes from pole conditions proceed exactly as in Sec.~\ref{subsec:dirac_scattering_bc} and Secs.~\ref{subsec:well_bound_state}--\ref{subsec:barrier_qnm}.

Specifically, for $V_{0}>0$ and $E>mc^{2}$, the effective potential is an attractive well that supports a finite number of bound
states. Meanwhile, for $V_{0}<0$ and $E>mc^{2}$, assuming $D(E)=1-8(E-mc^{2})|V_{0}|/(\hbar^{2}c^{2}\omega^{2})<0$ results in a barrier parameter $\lambda=-\tfrac{1}{2}\pm i\eta$, leading to the standard P\"oschl--Teller resonance pole families in the complex $k$-plane.

\section{Conclusions}
\label{sec:fr}

We investigated scattering in the one-dimensional Dirac equation with scalar and time-like vector
interactions by first mapping the relativistic problem, in a fully general way, into an equivalent
nonrelativistic Sturm--Liouville (Schr\"odinger-like) formulation for a single Dirac spinor component.
Under the spin-symmetry condition, the coupled first-order Dirac system decouples into a
second-order equation with an energy-dependent effective potential, so that standard scattering concepts
(transmission/reflection amplitudes and their analytic continuation) can be applied directly.
As an explicit application of this mapping, we then specialized to the P\"oschl--Teller profile, for which
supersymmetric quantum mechanics and shape invariance yield closed analytic expressions for the scattering
data and allow a unified discussion of continuum and discrete spectra.

Exploiting supersymmetric quantum mechanics and the shape-invariance property of the P\"oschl--Teller family,
we derived closed-form expressions for the transmission amplitude and the corresponding transmission probability. In the well configuration ($V_{0}>0$), the Dirac matching implies an energy-dependent
P\"oschl--Teller parameter $\lambda=\lambda(E)$, which controls the transparency through the factor
$\sin^{2}[\pi\lambda(E)]$; as a result, the reflection probability may be strongly suppressed when $\lambda(E)$
approaches integer values. The bound-state spectrum was recovered from the poles of the analytically continued
transmission amplitude, reproducing the known discrete levels and providing a unified description of continuum
scattering and bound states within the same analytic structure.

For the barrier configuration ($V_{0}<0$), we briefly discussed the pole structure associated with the analytic
continuation of the transmission amplitude. In the high-barrier regime, where the discriminant $D(E)$ is
negative and $\lambda=-\tfrac12\pm i\eta$, the poles form two symmetric families in the lower-half complex
$k$-plane, $k_{n}^{(\pm)}=\pm\omega\eta-i\omega(n+\tfrac12)$, illustrating the emergence of res\-o\-nance/quasi-normal
mode patterns for P\"oschl--Teller barriers. The transition at $D(E)=0$ provides a clear boundary between this
behavior and the low-barrier regime, in which the poles collapse to the imaginary axis and correspond to
antibound (virtual) states rather than resonances.

We also showed that the pseudospin-symmetry sector ($\Sigma=0$) can be obtained from the spin-symmetry
sector ($\Delta=0$) through the chiral transformation, which maps $\Sigma\leftrightarrow\Delta$,
$\phi_{\pm}\leftrightarrow\phi_{\mp}$ and $m\to -m$. This mapping allows the scattering results derived in one
sector to be translated to the other without repeating the full calculation.

Finally, we note that the P\"oschl--Teller barrier is widely used as an analytically tractable benchmark in
quasinormal-mode studies (see, e.g., \cite{CQG21:273:2003,PRD68:064007:2003,CQG26:163001:2009}), which further motivates the
pole-based discussion presented here. Possible extensions of the present analysis include a systematic study of resonant states and quasi-normal modes in the Dirac problem by solving the pole conditions self-consistently in the complex energy (or momentum)
plane, as well as the exploration of related shape-invariant interactions within the same SUSYQM framework.

\section*{Statements and Declarations}

\subsection*{Competing interests}
The authors declare no competing interests.

\subsection*{Data availability}
Data sharing is not applicable to this article as no datasets were generated or analyzed during the current study.

\begin{acknowledgements}
This work was supported in part by means of funds provided by CNPq, Brazil, Grant No. 308172/2023-0, FAPEMA and CAPES - Finance code 001.
\end{acknowledgements}

\appendix

\section{Derivation of $T(k)$ and $\mathcal{T}$}
\label{appendix:1}

In this Appendix, we provide a detailed derivation of Eqs.~\eqref{eq:T_gamma} and \eqref{eq:T_sinh_sin}. 

\subsection{Derivation of $T(k)$}
\label{appendix:1:1}

Starting from the recurrence relation given in Eq.~\eqref{eq:T_recursion}, we have
\begin{equation}
T(k;A)=\frac{A-ik}{-A-ik}\,T\!\left(k;A-\omega\right)\,.
\end{equation}
\noindent By iterating this relation $N$ times, the expression $T(k;A)$ becomes
\begin{equation}\label{eq:recursion_ext}
\begin{split}
T(k;A) ={} & \left(\frac{A-ik}{-A-ik}\right)\left(\frac{A-\omega-ik}{-A+\omega-ik}\right) \cdots \\
& \left(\frac{A-(N-1)\omega-ik}{-A+(N-1)\omega-ik}\right) T(k;A-N\omega) \,.
\end{split}
\end{equation}
\noindent By setting $A=N\omega$, $\lambda=A/\omega$ and $q=k/\omega$, and assuming the condition $T(k;0)=1$ for $\lambda\in\mathbb{R}$, Eq.~\eqref{eq:recursion_ext} takes the form
\begin{equation}\label{eq:recursion_ext2}
T(k)=\left(\frac{\lambda-iq}{-\lambda-iq}\right)\left(\frac{\lambda-1-iq}{-\lambda+1-iq}\right) \cdots \left(\frac{1-iq}{-1-iq}\right)\,.
\end{equation}
\noindent To express this product in terms of gamma functions, we identify the numerator and denominator as rising factorials, defined by the Pochhammer symbol \cite{FRANK2010}
\begin{equation}
(z)_{n}=\prod_{n=0}^{n-1}(z+n)=\frac{\Gamma(z+n)}{\Gamma(z)}\,.
\end{equation}
\noindent Specifically, the numerator can be written as:
\begin{equation}
(1-iq)_{n} = \prod_{n=0}^{\lambda-1}(1-iq+n) = \frac{\Gamma(\lambda+1-iq)}{\Gamma(1-iq)}\,,
\end{equation}
\noindent and the denominator as:
\begin{equation}
(-\lambda-iq)_{n} = \prod_{n=0}^{\lambda-1}(-\lambda-iq+n) =  \frac{\Gamma(-iq)}{\Gamma(-\lambda-iq)}\,.
\end{equation}
\noindent Substituting these results back into Eq.~\eqref{eq:recursion_ext2}, we arrive at the closed-form expression
\begin{equation}
T(k)=
\frac{\Gamma\left(\lambda+1-iq\right)\,
      \Gamma\left(-\lambda-iq\right)}
     {\Gamma\left(1-iq\right)\,
      \Gamma\left(-iq\right)}\,.
\end{equation}
\noindent This derivation confirms Eq.~\eqref{eq:T_gamma}, which is presented in Ref.~\cite{JPA21:L501:1988}.

\subsection{Derivation of $\mathcal{T}$}
\label{appendix:1:2}

To obtain the transmission probability $\mathcal{T}=\vert T(k)\vert^{2}$, we apply the Euler reflection formula $\Gamma(z)\Gamma(1-z)=\pi/\sin(\pi z)$ \cite{FRANK2010}, together with the following identities for the gamma function with imaginary arguments \cite{FRANK2010}:
\begin{eqnarray}
\vert\Gamma(iq)\vert^{2} &=& \frac{\pi}{q\sinh(\pi q)}\,,\\
\vert\Gamma(1+iq)\vert^{2} &=& q^{2}\vert\Gamma(iq)\vert^{2}=\frac{\pi q}{\sinh(\pi q)}\,.
\end{eqnarray}
\noindent These relations allow us to express the result in terms of trigonometric and hyperbolic function as
\begin{equation}\label{eq:T_squared_final}
\mathcal{T}=\frac{\sinh^{2}(\pi q)}{\sin^{2}(\pi\lambda)+\sinh^{2}(\pi q)}\,.
\end{equation}
\noindent By defining $p=\frac{\sinh(\pi q)}{\sin(\pi\lambda)}$, Eq.~\eqref{eq:T_squared_final} can be rewritten as 
\begin{equation}\label{eq:T_squared_final2}
\mathcal{T}=\frac{p^{2}}{1+p^{2}}\,.
\end{equation}
\noindent This derivation confirms Eqs.~\eqref{eq:T_sinh_sin}, in accordance with the standard result for P\"oschl--Teller potential \cite{FLUGGE1999}.

\bibliographystyle{spphys}       

\end{document}